\documentclass[twocolumn,superscriptaddress,amssymb,amsmath,aps,longbibliography,nofootinbib,prl,floatfix]{revtex4-2}
\usepackage{graphicx}
\usepackage{dcolumn}
\usepackage{bm}
\usepackage{float}
\usepackage{comment}
\usepackage{amsmath}
\usepackage{amssymb}
\usepackage{color}
\usepackage{subeqnarray}
\usepackage[utf8]{inputenc}
\usepackage[english]{babel}
\usepackage{mathrsfs}
\usepackage[bb=boondox]{mathalfa}
\usepackage{xcolor} 

\usepackage{amsbsy}
\usepackage{latexsym,epsfig,graphicx}
\usepackage{subfigure}
\usepackage{amsfonts}
\usepackage{amsmath}
\usepackage{xspace}
\usepackage{epstopdf}

\usepackage{array,booktabs}
\usepackage{babel}
\usepackage{multirow}
\usepackage[colorlinks=true, pdfstartview=FitV, linkcolor=red, citecolor=blue, urlcolor=blue]{hyperref}
\usepackage[normalem]{ulem}

\usepackage{cleveref,braket,xcolor}

\pdfoutput=1

\begin{document}

\title{Revealing Bosonic Exchange Symmetry in Two-Photon Temporal Wavefunction}

\author{Yefeng Mei}
\altaffiliation{These authors contributed equally to this work}
\affiliation{Department of Physics and Astronomy, Washington State University, Pullman, Washington 99164, USA}

\author{Yue Jiang}
\altaffiliation{These authors contributed equally to this work}
\affiliation{Department of Physics and Astronomy, University of Pittsburgh, Pittsburgh, Pennsylvania 15260, USA}

\author{Shengwang Du}
\email{dusw@purdue.edu}
\affiliation{Elmore Family School of Electrical and Computer Engineering, Purdue University, West Lafayette, Indiana 47907, USA}
\affiliation{Department of Physics and Astronomy, Purdue University, West Lafayette, Indiana 47907, USA}
\affiliation{Purdue Quantum Science and Engineering Institute, Purdue University, West Lafayette, Indiana 47907, USA}

\date{\today}

\begin{abstract}
The wavefunction of two identical bosons remains invariant under particle exchange—a fundamental quantum symmetry that underlies Bose-Einstein statistics. We report the direct experimental observation of bosonic exchange symmetry in the temporal wavefunction of photon pairs generated via spontaneous four-wave mixing in a three-level cold atomic ensemble. The measured two-photon temporal correlations show excellent agreement with theoretical predictions based on symmetrized bosonic wavefunctions. In addition, we perform time-resolved two-photon interference to reconstruct the complex temporal wavefunction. Both the amplitude and phase profiles exhibit clear symmetry under photon exchange, providing a direct confirmation of bosonic exchange symmetry in the time domain.
\end{abstract}

\maketitle

In quantum mechanics, exchange symmetry is a fundamental principle governing the behavior of identical particles that constitute our universe. For fermions—such as electrons—exchanging two particles introduces a negative sign in their joint quantum state, reflecting their antisymmetric wavefunction. In contrast, bosons obey a symmetric exchange rule: swapping their positions leaves the two-particle wavefunction unchanged. These contrasting symmetries give rise to two distinct quantum statistics: Fermi-Dirac statistics for fermions and Bose-Einstein statistics for bosons, which underpin the structure of matter and the behavior of many-body quantum systems. 

Recent advances in the control and manipulation of individual quantum systems have made it possible to probe exchange symmetry in few-particle quantum states. For example, the Pauli exclusion principle has been experimentally confirmed via spin blockade spectroscopy in a double quantum dot system~\cite{Ono2002}, where no two electrons can occupy the same quantum state. In the Hong-Ou-Mandel (HOM) interference experiment~\cite{HOM}, two indistinguishable photons incident on a 50:50 beam splitter always exit the same output port due to bosonic exchange symmetry. Such bosonic bunching (or fermionic antibunching) has also been observed in ultracold atom experiments~\cite{HBT-Nature-2007, KAUFMAN2018377}. Although exchange-phase measurements have been demonstrated with photons~\cite{Konrad-NP-2021, PhysRevApplied.18.064024, PhysRevLett.129.263602} and proposed for trapped atoms or ions~\cite{PhysRevLett.119.160401}, a direct measurement of the symmetric (or antisymmetric) space-time complex wavefunction of two identical bosons (or fermions) remains an outstanding challenge.

In this Letter, we report the direct observation of bosonic exchange symmetry in the temporal wavefunction of degenerate photon pairs generated from spontaneous four-wave mixing (SFWM) in a three-level laser-cooled $^{85}$Rb atomic ensemble. The theory derived from symmetrized bosonic wavefunctions show excellent agreement with two-photon coincidence measurements. To gain deeper insight, we further perform two-photon interference to reconstruct the full complex temporal wavefunction. The observed amplitude and phase are invariant under photon exchange, revealing a direct manifestation of bosonic symmetry in the structure of the two-photon wavefunction.

\begin{figure}[t]
\centering
\includegraphics[width=\linewidth]{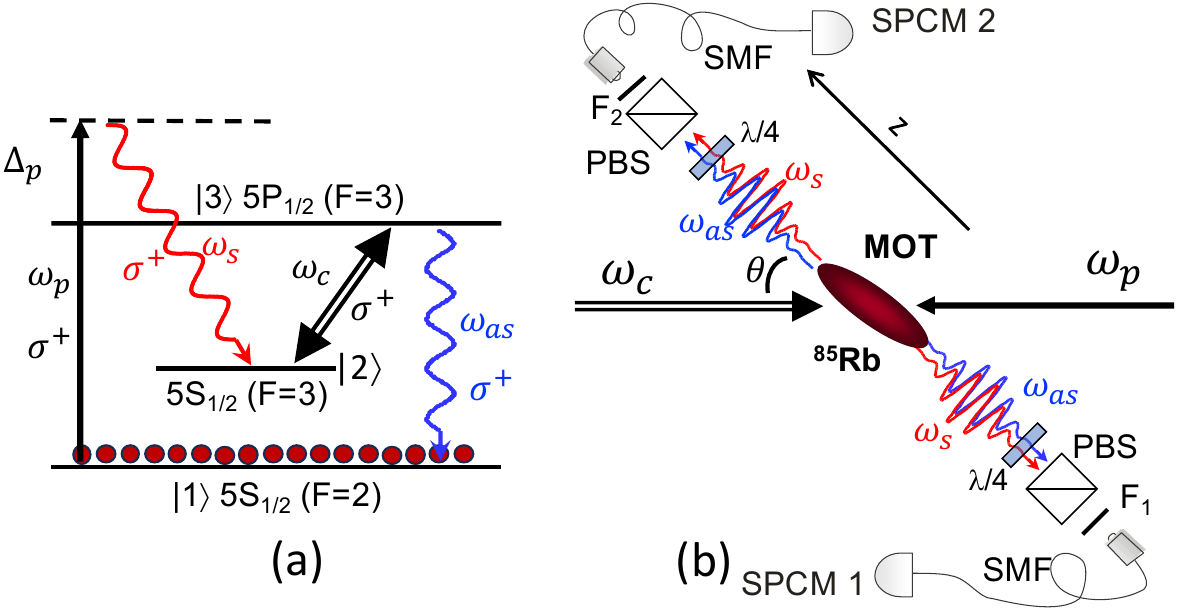}
\caption{Photon-pair generation via spontaneous four-wave mixing (SFWM) in a laser-cooled $^{85}$Rb atomic ensemble prepared in a magneto-optical trap (MOT). (a) Energy-level diagram of the three-level $^{85}$Rb atomic system. (b) Schematic of the SFWM optical setup. $\lambda/4$: quarter-wave plate; PBS: polarizing beam splitter; F$_{1,2}$: narrowband optical filters; SMF: single-mode fiber; SPCM: single-photon counting module.}
\label{fig:fig01}
\end{figure}

Figure~\ref{fig:fig01} presents our experimental platform for generating and measuring correlated photon pairs. Cold $^{85}$Rb atoms are loaded into a two-dimensional magneto-optical trap (MOT) with a length $L$= 1.5 cm and a temperature of about 10 $\mu$K ~\cite{2DMOT}. 
After the MOT loading time, the atoms are optically pumped to the ground state $|1\rangle$, as depicted in the energy level diagram in Fig.~\ref{fig:fig01}(a). SFWM \cite{Zhao:14, PhysRevLett.124.010509, Guo:17, PhysRevLett.100.183603} is driven by a pair of pump ($\omega_p$, 795 nm) and coupling ($\omega_c$, 795 nm) laser beams, counterpropagating at an angle $\theta=5^o$ with respect to the longitudinal $z$ axis, as illustrated in Fig.~\ref{fig:fig01}(b). The pump is blue-detuned by $\Delta_p$ from the $|1\rangle \rightarrow |3\rangle$ transition, while the coupling laser is resonant with the $|2\rangle \rightarrow |3\rangle$ transition. The spontaneously generated and phase-matched Stokes ($\omega_s$, $|3\rangle \rightarrow |2\rangle$) and anti-Stokes ($\omega_{as}$, $|3\rangle \rightarrow |1\rangle$) photons, moving backward along $z$ axis, are collected by two opposing single-mode fibers (SMFs) and detected by two single-photon counting modules (SPCMs 1 and 2: Excelitas SPCM-AQRH-16-FC). The two-photon coincidence is analyzed using a time-to-digital converter (Fast Comtec P7888). All optical fields are configured with the same circular polarization ($\sigma^+$). The hyperfine ground states $|1\rangle$ and $|2\rangle$ are separated by $\Delta_{12} = 2\pi \times 3.04$~GHz. Under phase-matching conditions, there are two equally weighted generation paths for each photon pair: (1) the Stokes photon is detected by SPCM 1 and the anti-Stokes photon by SPCM 2, and (2) the reverse detection configuration. By tuning the pump detuning to $\Delta_p = \Delta_{12}$, the Stokes and anti-Stokes photons become frequency degenerate ($\omega_{s0}=\omega_{as0}=\omega_0$) and thus indistinguishable, both experiencing the same linear propagation dispersion caused by electromagnetically induced transparency (EIT) \cite{EITPhysicsTodayHarris, EITRevModPhys}. As their spatial degrees of freedom are fixed - selected by the SMFs, in the following, we investigate the two-photon temporal wavefunction of these degenerate photon pairs.

Assuming the Stokes and anti-Stokes photons can be labeled according to their respective atomic transitions, the degenerate two-photon temporal wavefunction can be expressed as~\cite{Du:08}:
\begin{equation}
\Psi(t_s, t_{as}) = \langle 0| \hat{a}_{as}(t_{as}) \hat{a}_{s}(t_s) |\Psi\rangle = e^{-i\omega_0(t_s + t_{as})} \psi(\tau),
\label{eq:Psi_labeled}
\end{equation}
where $|\Psi\rangle$ is the two-photon state, $\hat{a}_{s}$ and $\hat{a}_{as}$ are the photon annihilation operators, $\omega_0$ is the central frequency, and $\tau = t_{as} - t_s$ is the relative delay between the photons arriving at their detectors. The relative two-photon temporal wavefunction $\psi(\tau)$ is given by
\begin{equation}
\psi(\tau) = \frac{L}{2\pi} \int d\varpi\, \kappa(\varpi) \Phi(\varpi) e^{-i\varpi \tau},
\label{eq:psi}
\end{equation}
where $\varpi = \omega_{as} - \omega_0$ is the frequency detuning of the anti-Stokes photon. The nonlinear coupling coefficient is
\begin{equation}
\kappa(\varpi) = \frac{-i \Omega_p \Omega_c \gamma_{13} \mathrm{OD} \mu_{32}/(2\mu_{13}L)}{(\Delta_p + i\gamma_{13}) \left[|\Omega_c|^2 - 4(\varpi + i\gamma_{13})(\varpi + i\gamma_{12})\right]},
\label{eq:kappa}
\end{equation}
where $\Omega_p$ and $\Omega_c$ are the Rabi frequencies of the pump and coupling lasers, respectively. OD is the optical depth for the $|1\rangle \rightarrow |3\rangle$ transition, $\mu_{mn}$ are dipole matrix elements, and $L$ is the medium length. $\gamma_{13}=2\pi\times 6$ MHz is the natural lifetime determined dephasing rate between $|1\rangle$ and $|3\rangle$. $\gamma_{12}=2\pi\times 0.025$ MHz is the ground-state dephasing rate between $|1\rangle$ and $|2\rangle$. The phase-matching function $\Phi(\varpi)$ is given by
\begin{equation}
\begin{split}
\Phi(\varpi) =& \mathrm{sinc} \left[ \frac{(k(\varpi) - k(-\varpi))L+(k_p-k_c)\cos{\theta}}{2} \right] \\
&\times e^{i [k(\varpi) + k(-\varpi)] L/2},
\end{split}
\label{eq:Phi}
\end{equation}
where the wavenumber $k(\varpi) = k_0 \sqrt{1 + \chi(\varpi)}$, with $k_0 = \omega_0/c$ and $c$ being the speed of light in vacuum. Here $k_p=\omega_p/c$ and $k_c=\omega_c/c$ are the pump and coupling laser wavenumbers, respectively.  The linear susceptibility under EIT conditions is
\begin{equation}
\chi(\varpi) = \frac{4\, \mathrm{OD}\, \gamma_{13}}{k_0 L} \frac{\varpi + i\gamma_{12}}{|\Omega_c|^2 - 4(\varpi + i\gamma_{13})(\varpi + i\gamma_{12})}.
\label{eq:chi}
\end{equation}

As discussed earlier, under perfect phase-matching, there exist two indistinguishable and equally probable photon-pair generation paths: one where the Stokes photon is detected by SPCM 1 and the anti-Stokes photon by SPCM 2, and the other with the reverse detection configuration. Because the photons are indistinguishable, it is fundamentally impossible to assign a Stokes or anti-Stokes label at the detectors. To account for exchange symmetry, we consider the symmetrized (bosonic) and antisymmetrized (fermionic) two-photon wavefunctions at the detectors, given by
\begin{equation}
\Psi_{\pm}(t_1, t_2) = e^{-i\omega_0(t_1 + t_2)} [\psi(\tau) \pm \psi(-\tau)],
\label{eq:Psi_sym}
\end{equation}
where $\tau = t_2 - t_1$, and the ``$+$'' and ``$-$'' signs correspond to bosonic and fermionic exchange symmetry, respectively. Let $\xi$ denote the duty cycle, $\eta$ the joint detection efficiency (including fiber coupling efficiency, filter transmissions, and detector efficiencies), $T$ the total data acquisition time, and $\delta t$ the time-bin width of the SPCMs. The two-photon coincidence counts are then given by
\begin{equation}
C_{\pm}(\tau) = \xi\eta T \delta t\, |\Psi_{\pm}(t_1, t_2)|^2 = \xi\eta T \delta t\, |\psi(\tau) \pm \psi(-\tau)|^2.
\label{eq:CC}
\end{equation}
In our experimental setup, the parameters are $\eta=0.05$, T=1200 s, $\delta t=2$ ns, with $\xi=1.5$\% for OD=7 and $\xi=1.2$\% for OD=25. In the following, we investigate how this temporal coincidence measurement reflects the bosonic nature of photons.

\begin{figure*}[t]
\centering
\includegraphics[width=\textwidth]{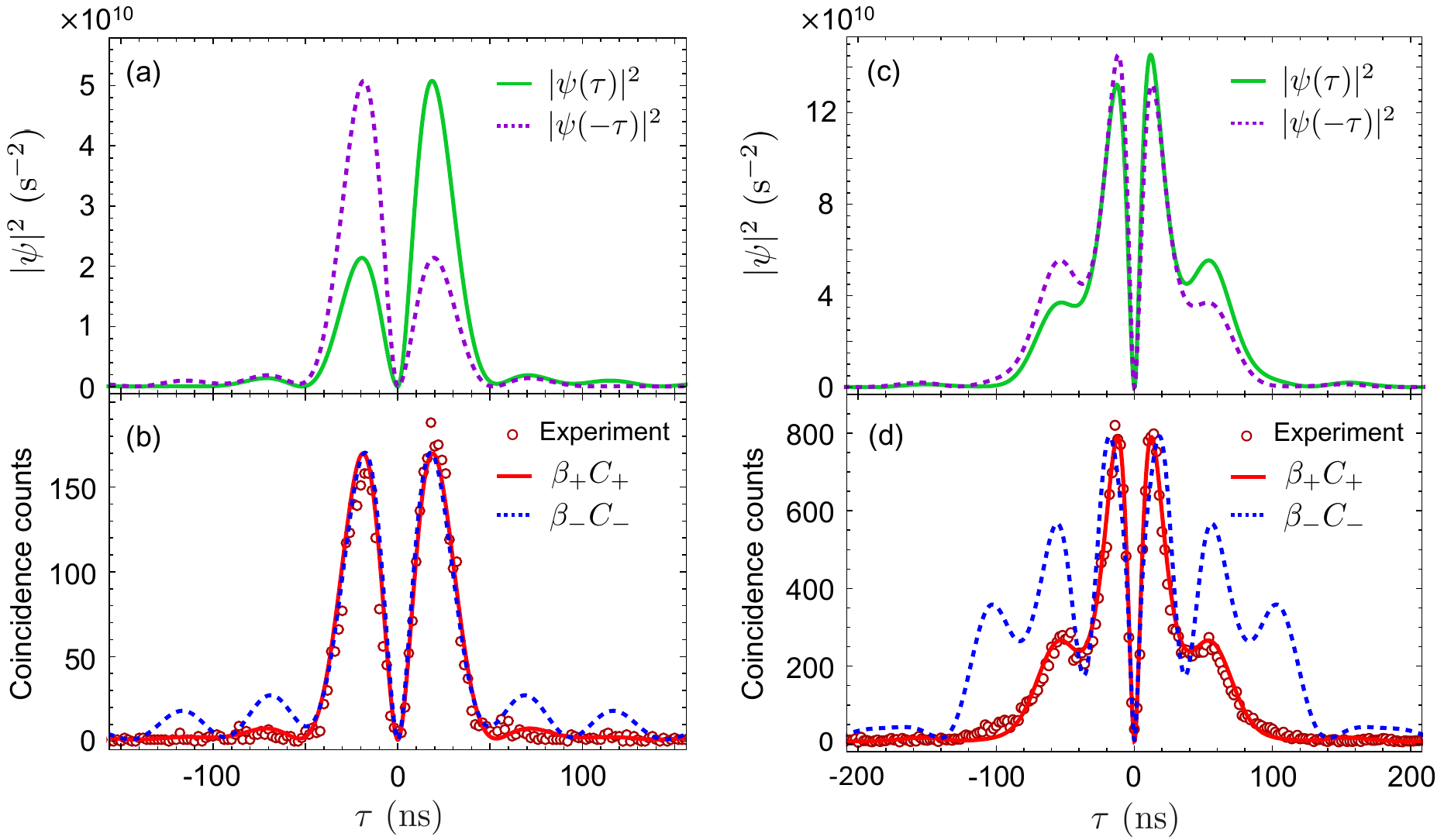}
\caption{Two-photon temporal wavefunctions at different optical depths (OD). (a), (b): OD = 7; (c), (d): OD = 25. (a), (c): Theoretical temporal wavefunctions $\psi(\tau)$ and $\psi(-\tau)$ of Stokes–anti-Stokes photon pairs. (b), (d): Measured two-photon coincidence counts (red circles), compared with theoretical predictions $\beta_{\pm} C_{\pm}(\tau)$ based on symmetrized (bosonic, $+$, red solid line) and antisymmetrized (fermionic, $-$, blue dashed line) wavefunctions. Other experiment parameters are $\Omega_p=2\pi\times 88.8$ MHz and $\Omega_c=2\pi\times 20.7$ MHz.}
\label{fig:fig02}
\end{figure*}


Figure \ref{fig:fig02} presents both theoretical and experimental results for different optical depths (OD). At OD = 7 [Fig.~\ref{fig:fig02}(a)], the temporal wavefunction $\psi(\tau)$ is neither symmetric [$\psi(\tau) = \psi(-\tau)$] nor antisymmetric [$\psi(\tau) = -\psi(-\tau)$], yet there exists substantial overlap between $\psi(\tau)$ and $\psi(-\tau)$. This partial overlap leads to pronounced interference effects in the two-photon correlations, depending on whether bosonic or fermionic exchange symmetry is applied, as described by Eqs.~\eqref{eq:Psi_sym} and~\eqref{eq:CC}. In Fig.~\ref{fig:fig02}(b), we compare the experimentally measured two-photon coincidence counts with theoretical predictions $\beta_{\pm} C_{\pm}(\tau)$, where $\beta_{\pm}$ are vertical scaling factors optimized to best fit the data. The symmetrized (bosonic) result, $\beta_{+} C_{+}(\tau)$, shows excellent agreement with the experimental measurements. In contrast, the antisymmetrized (fermionic) prediction, $\beta_{-} C_{-}(\tau)$, deviates significantly, particularly in the form of oscillatory features at $|\tau| >$ 50 ns, which are not observed experimentally. The fitted scaling factors are $\beta_{+} = 0.7$ and $\beta_{-} = 16.5$. We next increase the optical depth to OD = 25. As shown in Fig.~\ref{fig:fig02}(c), the overlap between $\psi(\tau)$ and $\psi(-\tau)$ increases, enhancing the sensitivity to exchange symmetry. Figure \ref{fig:fig02}(d) again compares the experimental data with theory, showing that the bosonic prediction $\beta_{+} C_{+}(\tau)$ (with $\beta_{+} = 1.0$) closely matches the measurement, while the fermionic counterpart $\beta_{-} C_{-}(\tau)$ (with $\beta_{-} = 179.5$) remains inconsistent. Both the correlation profiles and fitted scaling factors support the conclusion that the experimentally measured two-photon temporal wavefunctions obey bosonic exchange symmetry.

\begin{figure}[t]
\centering
\includegraphics[width=8.6 cm]{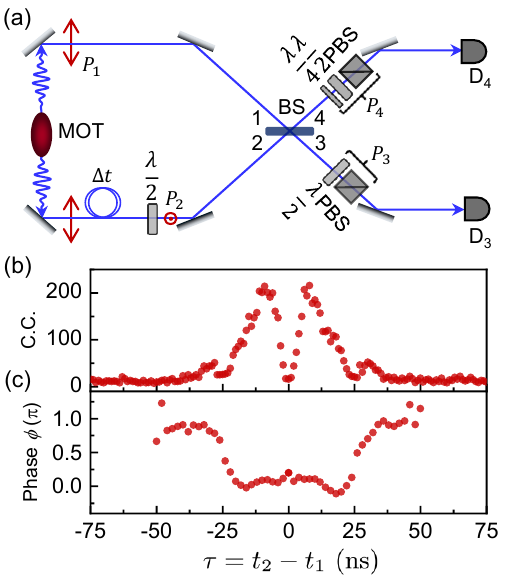}
\caption{Measurement of the complex temporal wavefunction of degenerate photon pairs using time-resolved two-photon interference. (a) Experimental setup for reconstructing the two-photon complex temporal wavefunction generated via backward SFWM. (b) Two-photon coincidence counts (C.C.) measured before the beam splitter (BS), representing the squared amplitude of the temporal wavefunction. (c) Reconstructed phase profile of the two-photon temporal wavefunction.}
\label{fig:fig03}
\end{figure}

To unambiguously verify bosonic exchange symmetry, we measure the phase profile of the two-photon temporal wavefunction using time-resolved two-photon interference~\cite{PhysRevLett.114.010401}. As illustrated in Fig.~\ref{fig:fig03}(a), the initially circularly polarized, counterpropagating photon pairs are converted into linearly polarized modes via quarter-wave ($\lambda/4$) plates and then coupled into polarization-maintaining (PM) SMFs. The fiber outputs are set to orthogonal polarizations: horizontal (P$_1$, $\leftrightarrow$) and vertical (P$_2$, $\updownarrow$), with P$_2$ rotated by a half-wave ($\lambda/2$) plate. The relative (optical) time delay $\Delta t$ between the two paths can be varied by setting the length difference of the PM SMFs. The photons are subsequently combined at a 50:50 beam splitter (BS), whose output ports (3 and 4) pass through polarizers P$_3$ and P$_4$ and are detected by two SPCMs (D$_3$ and D$_4$). Polarizer P$_3$ consists of a half-wave plate followed by a polarizing beam splitter (PBS), while P$_4$ includes a quarter-wave plate, a half-wave plate, and a PBS. Further experimental details are provided in Ref.~\cite{PhysRevLett.114.010401}. The amplitude profile of the two-photon wavefunction is obtained directly from the coincidence counts measured before the BS [Fig.~\ref{fig:fig03}(b)]. To reconstruct the phase, we collect twelve sets of coincidence measurements under different combinations of polarization projections (P$_3$ and P$_4$) at two distinct relative delays, $\Delta t = 1.0$ and $5.8$~ns. The extracted phase profile $\phi(\tau)$ is shown in Fig.~\ref{fig:fig03}(c). The observed symmetry in both the amplitude and phase functions confirms that the temporal wavefunction of the two indistinguishable photons exhibits bosonic exchange symmetry.

In summary, we have directly observed bosonic exchange symmetry in the temporal wavefunction of indistinguishable photon pairs generated via SFWM in a cold atomic ensemble. By measuring two-photon temporal correlations at varying optical depths, we confirmed that the experimental coincidence profiles agree with symmetrized theoretical predictions and are inconsistent with antisymmetric alternatives. Furthermore, through time-resolved two-photon interference, we reconstructed both the amplitude and phase of the complex temporal wavefunction and verified its exchange symmetry. These results provide a direct, time-domain confirmation of bosonic exchange symmetry and offer a new route toward probing fundamental quantum physics in few-particle systems.

S.D. acknowledges the support from NSF (Grants No. 2500662 and No. 2228725) and DOE (DE-SC0022069). Y.J. acknowledges the support from the PQI community award. Y.M. acknowledges the support from the WSU New Faculty Seed Grant.

\bibliography{biphoton}

\end{document}